
\input phyzzx
\input tables
\nopagenumbers
\voffset = -0.4in
\footline={\ifnum\pageno=1 \nulline \else\newfootline \fi}
\def\nulline{{\hfill}}
\def\newfootline{\advance\pageno by -1\hss\tenrm\folio\hss}
\def\N{{\cal N}}
\rightline {August 1993} \rightline {QMW--TH--93/22.}
\rightline {SUSX--TH--93/14.}
\title {String Loop Threshold Corrections \break For ${\bf Z}_N$
Coxeter Orbifolds.}
\author{D. Bailin$^{a}$, \ A. Love$^{b}$,  \ W.A.
Sabra$^{b}$\ and \ S. Thomas$^{c}$}
\address {$^{a}$School of Mathematical and Physical
Sciences,\break
University of Sussex, \break Brighton U.K.}
\address {$^{b}$Department of Physics,\break
Royal Holloway and Bedford New College,\break
University of London,\break
Egham, Surrey, U.K.}
\address {$^{c}$
Department of Physics,\break
Queen Mary and Westfield College,\break
University of London,\break
Mile End Road, London,  U.K.}
\abstract {
The moduli dependence of string loop threshold corrections to gauge coupling
constants is investigated for those ${\bf Z}_N$ Coxeter orbifolds with the
property that some twisted sectors have fixed planes for which the six-torus
${\bf T}_6$ can not be decomposed into a direct
sum ${\bf T}_4 \bigoplus{\bf T}_2$ with the fixed plane lying in ${\bf T}_2.$
\endpage

\REF\one {J. Ellis, S. Kelley and D. V. Nanopoulus, Phys. Lett. B260
(1991) 131; U. Amaldi, W. de Boer and H. Furstenau,  Phys. Lett. B260
(1991) 447.}
\REF\two{P. Ginsparg, Phys. Lett. B197 (1987) 139.}
\REF\three {V. S. Kaplunovsky, Nucl. Phys. B307 (1988) 145}
\REF\four {L. J.
Dixon, V. S. Kaplunovsky and J. Louis,  Nucl. Phys. B355 (1991) 649.}
\REF\five{L. J. Dixon, D. L\"{u}st and G. G. Ross, Phys. Lett. B272 (1991) 251;
L. E. Ibanez and D. L\"{u}st , Nucl. Phys. B382 (1992) 305..}
\REF\six{ D. Bailin and A. Love, Phys. Lett. B278 (1992) 125;
D. Bailin and A. Love, Phys. Lett. B292 (1992) 315.}
\REF\seven { I. Antoniadis, J. Ellis, R. Lacaze and D. V. Nanopoulus, Phys.
Lett. B268 (1991) 188;
 S. Kalara, J. L. Lopez and D. V. Nanopoulus, Phys. Lett. B269 (1991) 84.}
\REF\verlinde{ R. Dijkgraaf, E. Verlinde and H. Verlinde,
On Moduli Spaces of
Conformal Field
Theories with $c \geq 1$, Proceedings Copenhagen Conference,
Perspectives
in String Theory,
edited by P. Di Vecchia and J. L. Petersen,
World Scientific, Singapore, 1988.}
\REF\eight{ P. Mayr and S. Stieberger, preprint MPI-Ph/93-07, TUM-TH-152/93,
to appear in Nucl. Phys.}
\REF\nine{Y. Katsuki, Y. Kawamura, T. Kobayashi, N. Ohtsubo, Y. Ono and K.
Tanioka, Nucl. Phys. B341 (1990) 611;
T. Kobayashi and  N. Ohtsubo, preprint DPKU--9103.}
\REF\ten{M. Spalinski, Nucl. Phys. B377 (1992) 339.}
\REF\eleven{M. Spalinski,  Phys.  Lett. B275 (1992) 47;
J. Erler, D. Jungnickel and H. P. Nilles, Phys. Lett. B276 (1992) 303.}
\REF\twelve{D. Bailin, A. Love, W. A. Sabra and S. Thomas, preprint
QMW-TH-93/22, SUSX-TH-93/14.}

It is a remarkable feature of the extrapolation [\one] of the standard model
gauge
coupling constants using the renormalization group equations of the minimal
supersymmetric standard model that all three gauge coupling constants $g_3$,
$g_2$ and $g_1$ of $SU(3)\times SU(2)\times U(1)$  attain a common value at
about $10^{16}$ Gev.  At first sight, this is a problem for heterotic string
theory in view of the fact that the tree level gauge coupling constants of
string theory have a common value [\two, \three,  \four] at a string
unification
scale of about $10^{18}$ Gev.
However, it is possible in orbifold compactifications of string theory for the
string unification scale to be shifted down from $10^{18}$ to $10^{16}$ Gev by
moduli dependent string loop threshold corrections [\four, \five, \six],
whereas
in cases where there is no moduli dependence the effect is too small [\three,
\seven].

The moduli dependence of the threshold  corrections is associated [\four] with
orbifold twisted sectors $(h, g)$, where $h$ and $g$ refer to the twists in the
space and time directions, respectively, for which there is a complex plane
of the six torus
${\bf T}^6$ which is fixed by both $h$ and $g$. The calculation of this moduli
dependence is slightly easier when for all such fixed planes ${\bf T}^6$ can be
decomposed into a direct sum
${\bf T}^4\bigoplus{\bf T}^2$ with the fixed plane lying in ${\bf T}^2.$ In
this case a general result [\four] can be obtained. Calculations of the string
loop threshold corrections have also been carried out [\eight] for some
examples of ${\bf Z}_N$ Coxeter orbifolds for which this simplifying assumption
does not apply. We shall refer to these as non-decomposable orbifolds. It is
our purpose here to extend the calculations to all non-decomposable Coxeter
orbifolds.

The ${\bf Z}_N$ Coxeter orbifolds have been listed in ref. [\nine]. By
exploring the action of the point group on the lattice momentum and winding
numbers for all cases we have found that there is a quite small number of
examples which are non-decomposable in the sense described above. These are
displayed in table $1$.

The moduli-dependent string loop threshold corrections $\Delta_a$ to the gauge
coupling
constants $g^{-2}_a$ are determined by a subset of the twisted sectors $(h, g)$
of the orbifold. Only the twisted sectors for which there is a complex
plane for the six torus ${\bf T}^6$ fixed by both $h$ and $g$ contribute (the
sectors which are twisted sectors of an $\N=2$ space-time supersymmetric
theory.)
Explicitly [\four]
$$\Delta_a=\int_{\cal F}{d^2\tau\over \tau_2}\sum_{(h, g)}b_a^{(h,g)}{\cal
Z}_{(h, g)}(\tau, \bar\tau)-b_a^{\N=2}\int_{\cal F}{d^2\tau\over \tau_2},
\eqn\th$$
where ${\cal Z}_{(h, g)}$ denote the moduli dependent parts of
 the partition functions for the $\N=2$ twisted
sectors $(h, g)$,
$b_a^{(h,g)}$ is the contribution of the $(h, g)$ sector to the one-loop
renormalization group equations coefficient, $b_a^{\N=2}$ is the contribution
of
all $\N=2$ twisted sectors and $\cal F$ is the fundamental region for the world
sheet modular group $PSL(2,Z)$.
$${\cal F}=\left\{\tau : \tau_2>0, \quad \vert{\tau}\vert >1, \quad \vert
{\tau_1}\vert < {1\over2}  \right\},\quad \tau=\tau_1+i\tau_2.\eqn\nat$$

It is convenient [\eight] to rewrite \th\ in terms of a subset of $\N=2$
twisted
sectors $(h_0, g_0)$ referred to as the fundamental elements but with the
integration over an enlarged region
$\tilde{\cal F}$ depending on $(h_0, g_0)$. Then,
$$\Delta_a=\sum_{(h_0, g_0)}b_a^{(h_0,g_0)}\int_{\tilde{\cal F}}
{d^2\tau\over \tau_2}{\cal Z}_{(h_0, g_0)}(\tau, \bar\tau)-b_a^{\N=2}\int_{\cal
F}{d^2\tau\over \tau_2}.\eqn\tha$$
Here, the single twisted sector $(h_0, g_0)$ replaces a set of twisted sectors
which can be obtained from it by applying those $PSL(2, Z)$ transformations
which generate the fundamental region $\tilde{\cal F}$ of the world-sheet
modular symmetry group of ${\cal Z}_{(h_0, g_0)}$ from the fundamental region
${\cal F}$ of $PSL(2, Z)$. In particular, ${\cal Z}_{(h_0, g_0)}$ is invariant
under a congruence subgroup $\Gamma_0(n)$ of $PSL(2, Z)$ with $n=2$ or $3$,
where
$\Gamma_0(n)$  is defined to be the subgroup with
$$c= 0\   \hbox{(mod)} \  n \eqn\rain$$
of $PSL(2, Z)$ transformations
$$\tau\rightarrow {a\tau+b\over c\tau+d}\eqn\cult$$
where $a$, $b$, $c$, $d$ are integers with
$$ad-bc=1. \eqn\pink$$
When ${\cal Z}_{(h_0, g_0)}$ is invariant under $\Gamma_0(2)$ then
$$\tilde{\cal F}=\left\{ I, S, ST \right\} {\cal F},\eqn\ram$$
and when ${\cal Z}_{(h_0, g_0)}$ is invariant $\Gamma_0(3)$ then
$$\tilde{\cal F}=\left\{ I, S, ST, ST^2 \right\} {\cal F},\eqn\rama$$
where $S$ and $T$ denote the $PSL(2, Z)$ transformations
$$\eqalign{S:&\qquad \tau\rightarrow -{1\over\tau},\cr
T:&\qquad \tau \rightarrow \tau+1.}\eqn\sil$$

The calculation of the partition function terms ${\cal Z}_{(h_0, g_0)}$
proceeds as follows. For any particular orbifold with point group generated by
$\theta$, the action of $\theta$ on the basis vectors $e_a^i$ of the lattice of
the six torus is written in the form
$$\theta:\qquad e^i_a\rightarrow e^i_bQ_{ba}.\eqn\sus$$
If $n_a$ are the components of the winding number $2L^i$ with respect to this
basis
$$2L^i=\sum_{a} n_ae_a^i\eqn\aspen$$
and $m_a$ are the components of the lattice momentum $p_i$ with respect to the
dual basis $e^a_i$
$$p_i=\sum_{a} m_ae^a_i ,\eqn\pen$$
then it is convenient [\ten, \eleven] to write
$$w=\left (\matrix{
n_{1} \cr \vdots \cr n_6\cr}\right )\eqn\sun$$
and
$$p=\left (\matrix{
m_{1} \cr \vdots \cr m_6\cr}\right ).\eqn\moon$$
Then the action of $\theta$ on $w$ and $p$ is
$$\theta: \qquad w\rightarrow w'=Qw\eqn\no$$
and
$$\theta: \qquad p\rightarrow p'=(Q^t)^{-1}p.\eqn\no$$
Background metric and antisymmetric tensor fields $G$ and $B$ in the lattice
basis consistent with the action of the point group must satisfy
$$Q^tGQ=G,\qquad Q^tBQ=B\eqn\back$$
and the independent entries of $G$ and $B$ after satisfying \back\  are
the moduli (deformation parameters) for the orbifold.\footnote*{ We shall not
consider the inclusion of Wilson lines in the present treatment}

In the presence of the background fields the momenta $P_R$ and $P_L$ on the six
torus take the form
$$P_R={p\over2}-(G+B)w,\qquad P_L={p\over2}+(G-B)w.\eqn\nar$$
As a consequence of \nar\ the partition function ${\cal Z}_{(h_0,
g_0)}$
for a fundamental sector $(I, g_0)$ with no twist in the $\sigma $ direction
can always be written in the form [\eight]
$$\eqalign{{\cal Z}_{( I, g_0)}(\tau, \bar\tau)=&\sum_{(P_L, P_R)}q^{{1\over
2}P_L^t
G^{-1} P_L}
{\bar q}^{{1\over2} P_R^t G^{-1} P_R}\cr
&=\sum_{(p, w)} \textstyle{e^{2\pi i \tau_1p^tw}
e^{-\pi\tau_2({1\over2}p^tG^{-1}p-2p^tG^{-1}Bw+2w^t(G-BG^{-1}B)w)}}.}\eqn\deniro$$
where from the definition of ${\cal Z}_{(I, {g}_0)}$, only those $p$ and $w$
invariant
under the action of $g_0$ are summed over in eqn. (19).
 The fixed plane associated with  $g_0 =\theta^k$ has windings and momenta $w$
and $p$  determined by
$$Q^kw=w,\qquad {{({(Q^t)}^{-1})}}^kp=p\eqn\lor$$
and then $w$ and $p$ are each parametrized  by  two integers. Substituting the
form of $w$ and $p$
specified by \lor, introducing a metric $G_{\perp}$ and an
antisymmetric tensor $B_{\perp}$ for the two dimensional sub-lattice, and
defining standard moduli $T$ and $U$ by [\verlinde]
$$T=T_1+iT_2=2(B_{\perp}+i\sqrt{\det{G_{\perp}}} ),\quad
U=U_1+iU_2={1\over
{G_{\perp}}_{11}}({G_{\perp}}_{12}+i\sqrt{\det{G_{\perp}}}),\eqn\br$$
leads to an expression for  ${\cal Z}_{(I, g_0)}$ in terms of the components of
$w$ and $p$ in the fixed plane and the moduli $T$ and $U$.

In all cases, except for the ${\bf Z}_6$ orbifold with the $SU(3)\times SO(8)$
lattice, which is an example that has already been discussed in [\eight], we
find that the final expression for
${\cal Z}_{(I, g_0)}$ takes the form
$${\cal Z}_{( I, g_0)}=\sum_{{n_1,n_2}\atop {m_1,m_2}}
\textstyle{e^{2\pi i \tau(\gamma m_1n_1+\delta m_2n_2)}
e^{-\pi {\tau_2\over T_2U_2}\vert TUn_2+Tn_1-\gamma Um_1+\delta
m_2\vert^2},}\eqn\horror$$
where $n_1$, $n_2$, $m_1$ and $m_2$ are integers parametrizing the winding
number and momentum in the fixed plane.

Poisson resummation can be carried out on \horror, using the identity, for a
lattice $\Lambda$ and its dual lattice $\Lambda^*$,
$$\sum_{p\in\Lambda^*}
e^{[-\pi(p+\delta)^tC(p+\delta)+2\pi i p^t\phi)]}=V^{-1}_\Lambda {1\over
\sqrt{\det{C}}}
\sum_{l\in\Lambda}
e^{[-\pi(l+\phi)^tC^{-1}(l+\phi)-2\pi i \delta^t(l+\phi)]}\eqn\poisson$$
where $V^{-1}_\Lambda$ is the volume of the unit cell of the lattice $\Lambda,$
with the result
$$\tau_2 {\cal Z}_{(I, g_0)}={T_2\over \gamma\delta}\sum_{{n_1,n_2}\atop
{l_1,l_2}}
\textstyle{e^{2\pi i T {\det A}}e^{-\pi {T_2\over \tau_2U_2}\vert
\left (\matrix{1& U \cr}\right )A\left (\matrix{\tau\cr 1\cr }\right
)\vert^2}},\eqn\horr$$
where
$$A=\pmatrix{
n_1 &  {1\over\gamma}l_1 \cr
n_2 & {1\over\delta}l_2 \cr}\eqn\michael$$

The values for $\gamma$ and $\delta$ for the relevant twisted sectors of the
various orbifolds are given in table $2$.
For the ${\bf Z}_6-II-a$ orbifold and the  ${\bf Z}_{12}-I-a$ orbifold there is
no $U$ modulus associated with the fixed plane in the $(I, \theta^3)$ twisted
sector and in  \horror\ the modulus $U$ is replaced by the fixed numbers
$$U=U_0=-{1\over2} -i{\sqrt{3}\over2},\quad  \hbox {for}\  {\bf Z}_6-II-a,
\quad
U={\bar U}_0=-{1\over2} +i{\sqrt{3}\over2},\quad \hbox{for}\  {\bf Z}_{12}-I-a
.\eqn\wil$$

The $\tau$ integration in \tha\ can be carried out using the general approach
of refs. [\four, \eight].
The matrices $A$  in \horr\ are partitioned into orbits under the world sheet
modular symmetry group of ${\cal Z}_{(I, g_0)}$ which is in general a
congruence subgroup  $\Gamma_0(n)$ of  $PSL(2, Z)$ with $n=2$ or $3.$ The sum
over matrices $A$ (i.e., $l_1$, $l_2$, $n_1$ and $n_2$ )
is then replaced by the contribution of a representative matrix $A_0$ for each
orbit but with the region of $\tau$ integration enlarged to the union of all
regions $V\tilde{\cal F}$ of all matrices $V$ in $ \Gamma_0(n)$ that produce
distinct matrices $A=A_0V.$
In practice, it is slightly easier to work with integer-valued matrices
rather than the original matrices $A$.

We now complete the discussion of various non-decomposable orbifolds on a
case-by- case basis.
For ${\bf Z}_6-II-a,$ the fundamental sectors corresponding to the plane fixed
by $\theta^2$ may be taken to be
$(I, \theta^2)$, $(I, \theta^4).$
The partition functions ${\cal Z}_{(I, \theta^2)}$ and ${\cal Z}_{(I, \theta^4
)}$ turn out to be identical as can most easily be seen by applying the world
sheet modular transformation $ST^3S$ to ${\cal Z}_{(I, \theta^2)},$ remembering
that the modular transformation \cult\ has the action on the twisted sectors
$$(h, g) \rightarrow (h^dg^c, h^b g^a). \eqn\birgit$$
Thus we need only to double the contribution of ${\cal Z}_{(I, \theta^2)}.$
The partition function ${\cal Z}_{(I, \theta^2)}$ is invariant under the
congruence subgroup $\Gamma_0(3)$ of $PSL(2, Z)$  and the twisted sectors
$(\theta^2,I)$, $(\theta^2, \theta^2)$ and  $(\theta^2, \theta^4)$ are
obtained from it by the action of $S,$ $ST$ and $ST^2$, respectively.
The winding number and lattice momentum in the plane fixed by $\theta^2$ take
the form

$$w=\pmatrix{n_1\cr 0\cr n_1 \cr 0\cr n_1\cr n_2}, \qquad \qquad
p=\pmatrix{m_1\cr -m_1\cr m_1 \cr -m_1\cr m_1\cr m_2}.\eqn\hull$$
To evaluate the $\tau$ integration it is convenient to work in terms of the
matrices ${\tilde A}$ defined by
$$ A = \pmatrix{ {1\over 3}&0\cr 0&{1\over 3} }{\tilde A} \pmatrix
{3&0\cr0&1}, \quad \quad{\tilde A}=\pmatrix{
n_1 & l_1 \cr
n_2 & {\tilde l_2}\cr} \quad \hbox {with}\quad {\tilde l_2}=3l_2  \in 3Z
\eqn\chris$$
rather than the matrices $A$. The modular transformation \cult\  restricted to
$\Gamma_0(3)$ is equivalent to the transformation
$${\tilde A}\rightarrow {\tilde A}\pmatrix{
a &3b \cr
{c/3} &d\cr}\equiv {\tilde A}V\eqn\lolo$$
on the matrix ${\tilde A}$, where we must take
$$c=0 \ \hbox{(mod)}\ 3\eqn\joke$$
to stay within the $\Gamma_0(3)$ modular symmetry group of
${\cal Z}_{(I, \theta^2)}.$ The orbit containing the single matrix $A=0$ is
straightforward. The representative matrices for matrices with non-zero
determinant may be taken to be

$$\eqalign{{\tilde A}^I&=\pmatrix{k& j \cr 0 & p \cr},\quad
{\tilde A}^{II}={\tilde A}^I S=\pmatrix{j& -k\cr p & 0 \cr},\cr
{\tilde A}^{III}=&{\tilde A}^IST=\pmatrix{j&j -k\cr p & p \cr},\quad
{\tilde A}^{IV}={\tilde A}^IST^2=\pmatrix{j&2j -k\cr p & 2p \cr}},\eqn\mat$$
with $0\leq j<k$, $p\not=0$, and for
${\tilde A}^{I}$, ${\tilde A}^{III}$   and ${\tilde A}^{IV}$, $p=0 \
\hbox{(mod)} \ 3$.
The integration is over the double cover of the upper half plane. The matrices
with zero determinants may be represented by
$${\tilde A}^V=\pmatrix{0& j \cr 0 & p \cr},\quad
{\tilde A}^{VI}=\pmatrix{j& 0\cr p & 0 \cr},\eqn\al$$
with $j, p \in Z$, $(j, p) \not= (0, 0)$
and in ${\tilde A}^V,$  $p=0 \  \hbox{(mod)} \ 3$.
Because post-multiplying
${\tilde A}^V$ by the $\Gamma_0(3)$ matrices $V$ and  $T^n V$, $n$ an integer,
gives the same
result, to obtain the  ${\tilde A}^V$ contribution we must integrate over the
strip
$\left\{ \tau_2 >0, \quad \vert \tau_1\vert <{1\over2} \right\}.$ Also, because
post- multiplying
 ${\tilde A}^{VI}$ by the
$\Gamma_0(3)$ matrices  $S^t V S$ and  $S^tT^{3n} VS$ gives the same result, to
obtain the
${\tilde A}^{VI}$ contribution we must integrate over the strip
$\left\{ \tau'_2 >0, \quad \vert \tau'_1\vert <{3\over2} \right\}$ where
$\tau'=-\textstyle{{1\over\tau}}.$
We also need the value of $b_a^{\N=2}$ in \tha.
Performing $S$, $ST$ and $ST^2$ transformations on
${\cal Z}_{(I, \theta^2)}$ in turn to relate $b_{(\theta^2, I)}$
$b_{(\theta^2,\theta^2)}$ and $b_{(\theta^2,\theta^4 )}$ to
$b_{(I, \theta^2)}$ and including the equal contributions starting from ${\bf
Z}_{(1, \theta^4)}$ gives
$$b_a^{\N=2}=2b_a^{(I, \theta^2)}(1+{1\over3}+{1\over3}+{1\over3})=4
b_a^{(I, \theta^2)}\eqn\day$$
The final result for the contribution to $\Delta_a$ is
$$\eqalign{\Delta_a=&-2b_a^{(I, \theta^2)} \ln{\Big[kT_2\vert
\eta(T)\vert^4U_2\vert\eta(U)\vert^4\Big]}\cr
-& 2b_a^{(I,\theta^2)} \ln{\Big[kT_2\vert
\eta({T\over3})\vert^4U_2\vert\eta(3U)\vert^4\Big]}}\eqn\demons$$
where $\eta$ is the Dedekind function, and
$$k={8\pi\over 3{ \sqrt 3}}e^{1-\gamma_E}\eqn\euler$$
with $\gamma_E$ the Euler-Mascheroni constant.
Equation \demons\ displays the target space modular symmetry $\Gamma_T^0(3)
\times {(\Gamma_U)}_0(3),$ as a consequence of the transformation property
under modular transformations
$$T\rightarrow {aT+b\over cT+d}\eqn\green$$
of the Dedekind function
$$\vert \eta(T)\vert^4\rightarrow \vert \eta(T)\vert^4\vert
cT+d\vert^2.\eqn\schwarz$$

For the ${\bf Z}_6-II-a$ orbifold there is also a moduli dependent contribution
to $\Delta_a$ from the plane
fixed by $\theta^3$. In this case, the fundamental sector may be taken to be
$(I, \theta^3)$ and the partition function ${\bf Z}_{(I, \theta^3)}$ is
invariant under $\Gamma_0(2)$ of $PSL(2, Z)$. The winding number and lattice
momentum in the plane fixed by $\theta^3$ takes the form
$$w=\pmatrix{n_1\cr n_2\cr 0 \cr n_1\cr n_2\cr 0}, \qquad \qquad
p=\pmatrix{m_1\cr  m_2\cr -m_1-m_2 \cr m_1\cr m_2\cr 0},\eqn\contact$$

and, in terms  of $m_1$, $m_2$, $n_1$ and $n_2$,
${\bf Z}_{(I, \theta^3)}$ here has the same form as
for the ${\bf Z}_4-a$
orbifold example considered in [\eight]. The calculation is in all respects
similar to that case except that $U$ now takes the fixed value $U_0$ of \wil.
Thus, the contribution to the string loop threshold correction is
$$\eqalign{\Delta_a=&-b_a^{(I, \theta^3)} \ln{\Big[k{{\tilde T}_2\over4}\vert
\eta({{\tilde T}\over2})\vert^4
(-{\sqrt3\over2})\vert\eta(U_0)\vert^4\Big]}\cr
-& {1\over 2}b_a^{(I, \theta^3)} \ln{\Big[k{\tilde T}_2\vert \eta({{\tilde
T}\over2})\vert^4
(-{\sqrt3\over2})\vert\eta(U_0)\vert^4\Big ]}}\eqn\devil$$
where we have denoted the modular parameter associated with the plane fixed by
$\theta^3$ by ${\tilde T}$. This contribution to $\Delta_a$ has the target
space modular symmetry
$\Gamma_{\tilde T}^0(2)$.

In the case of ${\bf Z}_6-II-c$ orbifold the winding number and momentum for
the plane fixed by $\theta^2$ takes the form
$$w=\pmatrix{0\cr 0\cr n_1 \cr 0\cr n_1\cr n_2}, \qquad \qquad
p=\pmatrix{0\cr  0\cr 2m_1 \cr -2m_1\cr m_1\cr m_2},\eqn\pinhead$$
and, in terms of $n_1$, $n_2$, $m_1$ and $m_2$,
${\bf Z}_{(I, \theta^2)}$ takes the same form as ${\bf Z}_{(I, \theta^2)}$ for
the
${\bf Z}_6-II-a$ orbifold. The calculation is in all respects similar leading
to the contribution \demons\ to $\Delta_a$. For the plane fixed by $\theta^3$
the $SU(3)\times SO(7)\times SU(2)$ lattice is decomposable. Consequently the
general result of ref [\four] applies, and we get the contribution to the
string loop threshold correction
$$\Delta_a= -{\hat b}_a\ln \Big[ k{\tilde T}_2\vert\eta({\tilde
T})\vert^4{\tilde U}_2\vert\eta({\tilde U})\vert^4\Big ]\eqn\arafat$$
with full $PSL_{\tilde T}(2, Z)\times PSL_{\tilde U}(2, Z)$ modular symmetry,
where
${\hat b}_a$ is the contribution to $b_a^{\N=2}$ from the $(I, \theta^3)$ , $(
\theta^3, I )$ and $(\theta^3, \theta^3)$ sectors.

For the ${\bf Z}_{12}-I-a$ orbifold there are $5$ fundamental sectors which may
be taken to be
$(I, \theta^3)$ , $(I, \theta^9)$, $(\theta^6, \theta^3)$,
$(\theta^6,  \theta^9)$ and $(I, \theta^6),$ with the other twisted sectors
being generated from these by the action of $S$ and $ST$. The partition
functions ${\bf Z}_{(I, \theta^9)}$, ${\bf Z}_{( \theta^6, \theta^3)}$ and
${\bf Z}_{(\theta^6, \theta^9)}$ are identical to
${\bf Z}_{(I, \theta^3)}$as can be seen by applying modular transformations to
${\bf Z}_{(I, \theta^3)},$ and it can be shown by direct calculation that
${\bf Z}_{(I, \theta^6)}$ is also identical to
${\bf Z}_{(I, \theta^3)}.$ Thus, the complete result is identical to
$5{\bf Z}_{(I, \theta^3)}.$ The winding number and momentum for the plane fixed
by $\theta^3$ takes the form
$$w=\pmatrix{n_1\cr n_2\cr 0\cr n_1\cr n_2\cr 0}, \qquad \qquad
p=\pmatrix{m_1\cr  -m_1\cr m_1 \cr 0\cr 0\cr m_2},\eqn\hellraiser$$
and in terms of the integers $n_1$, $n_2$, $m_1$ and $m_2$,
${\bf Z}_{(I, \theta^3)}$ has the same form as ${\bf Z}_{(I, \theta^3)}$ for
${\bf Z}_6-II-a$ with $U=U_0$ replaced by $U={\bar U}_0$ as in \wil.
Thus the string loop threshold correction is
$$\eqalign{\Delta_a=&-5b_a^{(I, \theta^3)} \ln{\Big[k{{ T}_2\over4}\vert
\eta({{T}\over2}\Big )\vert^4
({\sqrt3\over2})\vert\eta({\bar U}_0)\vert^4\Big ]}\cr
-&{5 \over 2}b_a^{(I, \theta^3)} \ln{\Big [k{ T}_2\vert \eta({{
T}\over2})\vert^4
({\sqrt3\over2}) \vert\eta({\bar U}_0)\vert^4\Big ]}}\eqn\israel$$
with the symmetry group $\Gamma_{T}^0(2)$.

Of the cases not considered in [\eight], it remains to discuss ${\bf Z}_4-b$.
For this orbifold there is a single fundamental sector which may be taken to be
$(I, \theta^2)$ with $(\theta^2, I)$ and $(\theta^2, \theta^2)$ obtained from
it by $S$ and $ST$ transformations. The winding number and momentum in the
plane fixed by $\theta^2$ takes the form

$$w=\pmatrix{n_1\cr 0\cr n_1\cr 0\cr 0\cr n_2}, \qquad \qquad
p=\pmatrix{m_1\cr  -m_1\cr m_1 \cr 0\cr 0\cr m_2},\eqn\hellbound$$
and the partition function
${\bf Z}_{(I, \theta^2)}$ is invariant under the congruence subgroup
$\Gamma_0(2)$ of $PSL(2, Z)$. The situation somewhat resembles that for the
$(I, \theta^2)$ sector of the ${\bf Z}_6-II-a$ orbifold except that in \chris\
we have ${\tilde l}_2=2l_2 \in 2Z$, in \mat\ only the first three matrices are
required with $p$ even in both ${\tilde A}^I$and ${\tilde A}^{III}$, and for
${\tilde A}^{V}$ of \al\ one
must also take $p$ even. The range of integration of ${\tilde A}^V$ is modified
to
$$\left\{ \tau'_2 >0,\qquad \vert \tau'_1\vert<1  \right\}\eqn\jose$$
where $\tau'=-{1\over \tau}$.
Also, we now have
$$b_a^{\N=2}=(1+{1\over2}+{1\over2} )b_a^{(I, \theta^2)}=2b_a^{(I, \theta^2
)}.\eqn\bond$$
The final result for $\Delta_a$ is
$$\eqalign{\Delta_a=&-b_a^{(I, \theta^2)} \ln {\Big(kT_2\vert
\eta(T)\vert^4 U_2\vert\eta(2U)\vert^4\Big )}\cr
-& b_a^{(I, \theta^2)} \ln{\Big (kT_2\vert
\eta({T\over2})\vert^4U_2\vert\eta(U)\vert^4\Big)}}\eqn\jordan$$
which displays target space modular symmetry $\Gamma_T^0(2)\times
{(\Gamma_0)}_U(2)$.

The string loop threshold corrections and their target space modular symmetries
are summarized in table $3$.

In conclusion, the string loop threshold corrections for gauge coupling
constants have been obtained for all non-decomposable ${\bf Z}_N$ Coxeter
orbifolds, i.e., where some twisted sectors have fixed planes for which the
six-torus ${\bf T}^6$
can not be decomposed into a direct sum
 ${\bf T}^4 \oplus{\bf T}^2$ with the fixed plane lying in ${\bf T}^2.$

The target space modular symmetry group associated with such fixed planes is
always a  congruence subgroup $\Gamma^0(2)$ or  $\Gamma^0(3)$ of the full
$PSL(2, Z)$ for the $T$ moduli and this is also the case for $U$ moduli
in relation to the subgroups $\Gamma_0(2)$ or  $\Gamma_0(3)$, with one
exception. These modular symmetry groups are in agreement with those obtained
by studying the spectrum of states for these orbifolds [\twelve].
\vskip 1cm
\centerline{ ACKNOWLEDGEMENTS}
This work was supported in part by S.E.R.C. and the Royal Society.
\vfill\eject
\centerline{\underbar{ TABLE CAPTIONS}}
\underbar {TABLE 1}: \hfill\break

Non decomposable ${\bf Z}_N$ orbifolds. For the point group generator $\theta,$
we display $(\xi_1, \xi_2, \xi_3)$ such that the action of $\theta$ in the
complex orthogonal basis is \break
$(e^{2\pi i\xi_1}, e^{2\pi i\xi_2}, e^{2\pi i\xi_3})$.
\vskip 1cm
\underbar {TABLE 2}:\hfill\break

Values of parameters $\gamma$ and $\delta$ for relevant twisted sectors $(h_0,
g_0)$ of the non-decomposable ${\bf Z}_N$ orbifolds.
\vskip 1cm
\underbar {TABLE 3}:\hfill\break

String loop threshold corrections $\Delta_a$ and target space modular symmetry
groups for the non-decomposable ${\bf Z}_N$ orbifolds. The constants $k$, $U_0$
and ${\bar U}_0$ are given in \euler, \wil..
\vfil\eject
\centerline {TABLE 1}
\vskip 0.5cm
\begintable
Orbifold | Point Group Generator $\theta$|Lattice\cr
$Z_4-a$| $(1,1,-2)/4$| $SU(4)\times SU(4)$\cr
$Z_4-b$| $(1,1,-2)/4$|$SU(4)\times SO(5)\times SU(2)$\cr
$Z_6-II-a $|$(2,1,-3)/6$|$SU(6)\times SU(2)$\cr
$Z_6-II-b $|$(2,1,-3)/6$|$SU(3)\times SO(8)$\cr
$Z_6-II -c$|$(2,1,-3)/6$|$SU(3)\times SO(7)\times
SU(2).$\cr
$Z_8-II -a$|$(1,3,-4)/8$|$SU(2)\times SO(10)$\cr
$Z_{12}-I-a$| $(1,-5,4)/12$|$E_6$
\endtable
\vskip 1.5cm
\centerline {TABLE 2}
\vskip 0.5cm
\begintable
Orbifold | Twisted Sector |$\gamma$|$\delta$\cr
$Z_4-a$ | $(I, \theta^2)$ |2|2\cr
$Z_4-b$ | $(I, \theta^2)$ |2|1\cr
$Z_6-II-a$ | $(I, \theta^3)$ |2|2\nr
| $(I, \theta^2)$ |3|1\cr
$Z_6-II-c$ | $(I, \theta^2)$ |3|1\cr
$Z_8-II-a$ | $(I, \theta^2)$ |2|1\cr
$Z_{12}-I-a$ | $(I, \theta^3)$ |2|2
\endtable
\vfill\eject
\centerline {TABLE 3}
\vskip 0.5cm
\begintable
Orbifold | $\Delta_a$ |Target Space Duality Symmetry\cr
${\bf Z}_4-a$ | $-b_a^{(I, \theta^2)}\ln
\Big(k{T_2\over4}\vert\eta({T\over2})\vert^4U_2\vert\eta(U)\vert^4\Big)$|
$\Gamma_T^0(2)\times PSL_U(2, Z)$\nr
|  $-{1\over2}b_a^{(I, \theta^2)}\ln
\Big(kT_2\vert\eta({T\over2})\vert^4U_2\vert\eta(U)\vert^4\Big)$ |   \cr
${\bf Z}_4-b$ | $-b_a^{(I, \theta^2)}\ln
\Big(kT_2\vert\eta(T)\vert^4U_2\vert\eta(2U)\vert^4\Big)$|
$\Gamma_T^0(2)\times{(\Gamma_U)}_0(2)$\nr
|  $-b_a^{(I, \theta^2)}\ln
\Big(kT_2\vert\eta({T\over2})\vert^4U_2\vert\eta(U)\vert^4\Big)$ |   \cr
${\bf Z}_6-II-a$ | $-2b_a^{(I, \theta^2)}\ln
\Big(kT_2\vert\eta(T)\vert^4U_2\vert\eta(U)\vert^4\Big)$|
$\Gamma_T^0(3)\times{(\Gamma_U)}_0(3)$\nr
|  $-2b_a^{(I, \theta^2)}\ln
\Big(kT_2\vert\eta({T\over3})\vert^4U_2\vert\eta(3U)\vert^4\Big)$ |   \nr
 | $-{1\over 2}b_a^{(I, \theta^3)}\ln \Big(k{\tilde T_2}\vert\eta({\tilde
T\over2})\vert^4{(-{\sqrt 3\over2})}\vert\eta(U_0)\vert^4\Big)$|
$\Gamma_{\tilde T}^0(2)$\nr
| $-b_a^{(I, \theta^3)}\ln \Big(k{{\tilde T_2}\over4}\vert\eta({\tilde
T\over2})\vert^4{(-{\sqrt 3\over2})}\vert\eta(U_0)\vert^4\Big)$|   \cr
${\bf Z}_6-II-b$ | $-2b_a^{(I, \theta^2)}\ln
\Big(k{T_2\over3}\vert\eta({T\over3})\vert^4{U_2\over3}\vert\eta({U+2\over3})\vert^4\Big)$| $\Gamma_T^0(3)\times\Gamma_{U+2}^0(3)$\nr
|  $-2b_a^{(I, \theta^2)}\ln
\Big(kT_2\vert\eta(T)\vert^4U_2\vert\eta(U+2)\vert^4\Big)$ |   \nr
 | $-{\hat b}_a\ln (k{\hat T}_2\vert\eta({\hat T})\vert^4{\hat
U}_2\vert\eta({\hat U})\vert^4\Big)$| $PSL_{\hat T}(2, Z)\times PSL_{\hat U}(2,
Z)$ \cr
${\bf Z}_6-II-c$ | $-2b_a^{(I, \theta^2)}\ln
\Big(kT_2\vert\eta(T)\vert^4U_2\vert\eta(U)\vert^4\Big)$|
$\Gamma_T^0(3)\times{(\Gamma_U)}_0(3)$\nr
|  $-2b_a^{(I, \theta^2)}\ln
\Big(kT_2\vert\eta({T\over3})\vert^4U_2\vert\eta(3U)\vert^4\Big)$ |   \nr
 | $-{\hat b}_a\ln \Big(k{\hat T}_2\vert\eta({\hat T})\vert^4{\hat
U}_2\vert\eta({\hat U})\vert^4\Big)$| $PSL_{\hat T}(2, Z)\times PSL_{\hat U}(2,
Z)$ \cr
${\bf Z}_8-II-a$ | $-5b_a^{(I, \theta^2)}\ln
\Big(kT_2\vert\eta(T)\vert^4U_2\vert\eta(2U)\vert^4\Big)$| $\Gamma_T^0(2)\times
{(\Gamma_U)}_0(2)$\nr
|  $-5b_a^{(I, \theta^2)}\ln
\Big(kT_2\vert\eta({T\over2})\vert^4U_2\vert\eta(U)\vert^4\Big)$ |   \cr
${\bf Z}_{12}-I-a$ | $-5b_a^{(I, \theta^3)}\ln
\Big(k{T_2\over4}\vert\eta({T\over 2})\vert^4({\sqrt 3\over2})\vert\eta(\bar
U_0)\vert^4\Big)$| $\Gamma_T^0(2)$\nr
|  $-{5\over2}   b_a^{(I, \theta^3)}\ln
\Big(kT_2\vert\eta({T\over2})\vert^4({\sqrt
3\over2})\vert\eta(\bar U_0)\vert^4\Big)$ |
\endtable
\vfill\eject
\refout
\end